\newif\ifusingllncs

\ifusingllncs 
\documentclass[10pt,draft]{llncs}
\author{Norman Danner\inst{1} \and James S.~Royer\inst{2}}
\institute{%
Department of Mathematics and Computer Science, Wesleyan University,
Middletown, CT 06459, USA; Email: \email{ndanner@wesleyan.edu} \and
Department of Electrical Engineering and Computer Science, Syracuse University,
Syracuse, NY 13210, USA; Email: \email{royer@ecs.syr.edu}}

\else 
\documentclass[11pt,final]{ndrpt}
\author{Norman Danner and James S. Royer}
\address{Department of Mathematics and Computer Science, Wesleyan University,
Middletown, CT 06459, USA}
\email{ndanner@wesleyan.edu}
\address{%
Department of Electrical Engineering and Computer Science, Syracuse University,
Syracuse, NY 13210, USA}
\email{royer@ecs.syr.edu}
\thanks{\textbf{This paper is to be first published in S.B.~Cooper, B.L\"owe, and A.~Sorbi (eds.), \emph{Computation in the Real World (Proceedings Computability in Europe, 2007, Sienna)}, vol.~4497 of \emph{Lecture Notes in Computer Science}, Springer-Verlag, Berlin, 2007.}}
\begingroup
\catcode`$=9
\draftfoot{DRAFT--Please do not redistribute\hfill \hbox{$Revision: 1.130.2.1 $, $Date: 2007-03-19 21:50:24 $}}
\endgroup
\fi

\title{Time-complexity semantics for feasible affine recursions}

\ifusingllncs\usepackage{amsmath}\fi
\usepackage{amssymb}
\usepackage{bussproofs}
\EnableBpAbbreviations
\def\LeftLabelBold#1{\LeftLabel{\textbf{#1}}}
\def\InfRule#1{\textbf{#1}}
\usepackage[final]{listings}
\usepackage[numbers]{natbib}
\ifusingllncs\else \usepackage{ndthms}\fi
\usepackage{suffix}

\let\union=\cup
\let\cross=\times
\def\dom{\mathrm{Dom}\;}
\def\id{\mathop{\mathrm{id}}\nolimits}
\def\setseparator{\mid}
\newcommand{\set}[2][\relax]{
  \ifx#1\relax
    \{#2\}
  \else
    \ifx#1\left
      \left\{#2\right\}
    \else
      \csname #1l\endcsname\{#2\csname #1r\endcsname\}
    \fi
  \fi
}
\newcommand{\setst}[3][\relax]{\set[#1]{#2\setseparator#3}}

\def\arrow{\mathbin{\rightarrow}}
\def\eqdef{=_{\mathrm{df}}}
\def\llambda{{\lambda\hskip-.45em\lambda}}	
\def\fv{\mathop{\mathrm{fv}}\nolimits}
\def\oftype{\mathbin{:}}
\def\tmden#1{{\lbrack\!\lbrack{#1}\rbrack\!\rbrack}}

\def\ATR{\mathsf{ATR}}

\def\Nat{\mathsf{N}}
\def\Tally{\mathsf{T}}

\def\typing#1#2#3#4{#1;#2\proves#3\oftype#4}
\def\ityping#1#2#3{\typing{#1}{\underline{~}}{#2}{#3}}
\def\GDtyping#1#2{\typing\Gamma\Delta{#1}{#2}}
\def\Gtyping#1#2{\ityping\Gamma{#1}{#2}}

\def\tctyping#1#2#3{#1\proves#2\oftype#3}
\def\Stctyping#1#2{\tctyping\Sigma{#1}{#2}}

\def\semanticOp#1{\mathop{\smash{\mathit{#1}}}\nolimits}
\let\combfont\mathsf
\def\comb#1{\mathop{\smash{\combfont{#1}}}\nolimits}

\def\afflambda{\lambda_r}
\def\cons{\comb{c}}
\def\destr{\comb{d}}
\def\test{\comb{t}}
\def\cond#1#2#3{\mbox{$\combfont{if}~#1~\combfont{then}~#2~\combfont{else}~#3$}}
\def\down{\comb{down}}
\def\crec{\comb{crec}}
\def\rec{\comb{rec}}

\let\apprby\sqsubseteq
\def\apprbypot{\apprby_{\mathrm{pot}}}

\let\base\mathsf
\let\bmax\vee
\def\bz{\mathbf{0}}
\def\bone{\mathbf{1}}

\def\cl#1#2{#1#2}
\WithSuffix\def\cl*#1#2{(#1)#2}
\def\cost{\semanticOp{cost}}

\def\dally{\semanticOp{dally}}
\let\dmnd\Diamond

\def\Env#1{\text{$#1$-$\mathrm{Env}$}}
\def\emptyenv{[]}
\let\eps\varepsilon
\let\evalto\downarrow
\def\extend#1#2#3{#1[#2 \mapsto #3]}

\def\lh#1{|#1|}

\let\plusmax\uplus
\def\pmj{\odot}
\def\pmjb{\pmj_{\base b}}
\def\pot{\semanticOp{pot}}
\def\potden#1{\langle\!\langle#1\rangle\!\rangle}
\let\proves\vdash

\let\shiftsto\propto
\def\stdcrec{\crec\,a\,(\afflambda f.\lambda\vec v.t)}
\def\strictsubtype{\mathrel{<:}}

\def\subtype{\mathrel{\leq:}}

\def\tail{\semanticOp{tail}}
\def\tcden#1{{\|#1\|}}

\def\val{\semanticOp{val}}

\ifusingllncs 
\spnewtheorem{thm}{Theorem}{\normalfont\scshape}{\normalfont\slshape}
\spnewtheorem{lem}[thm]{Lemma}{\normalfont\scshape}{\normalfont\slshape}
\spnewtheorem{prop}[thm]{Proposition}{\normalfont\scshape}{\normalfont\slshape}
\spnewtheorem{cor}[thm]{Corollary}{\normalfont\scshape}{\normalfont\slshape}
\spnewtheorem*{defn}{Definition}{\normalfont\scshape}{\normalfont}
\else
\newenvironment{lem}{\begin{lemma}}{\end{lemma}}
\fi

\ifusingllncs\else
\makeatletter
\renewcommand{\paragraph}{\@startsection%
	{paragraph}%
	{4}%
	{0in}%
	{.5\baselineskip}%
	{-.5em}%
	{\itshape}%
}
\makeatother
\fi

\let\pgmfont\mathsf

\def\ATS{\textit{ATS}}

\let\cat\oplus

\def\level{\semanticOp{level}}
\let\lollipop\multimap

\def\PCF{\mathsf{PCF}}
\def\pad{\semanticOp{pad}}
\let\phi\varphi
\def\pgm#1{\mathop{\pgmfont{#1}}\nolimits}

\let\restr\upharpoonright

\def\taillh{\semanticOp{tail\_len}}

\ifusingllncs\else\usepackage[final]{hyperref}\fi

\begin{document}

\maketitle

\begin{abstract} 
  The authors' $\ATR$ programming formalism is a version of
  call-by-value $\PCF$ under a complexity-theoretically motivated type
  system.  
  $\ATR$ programs run in type-$2$ polynomial-time and all
  standard type-$2$ basic feasible functionals are
  $\ATR$-definable
  ($\ATR$ types are confined to levels  $0$, $1$, and~$2$).
  A limitation of the original version of $\ATR$ is that
  the only directly expressible recursions are
  tail-recursions.  Here we extend $\ATR$ so that a broad range
  of affine recursions are directly expressible.  In particular,
  the revised $\ATR$ can fairly naturally express
  the classic insertion- and selection-sort algorithms,
  thus overcoming a sticking point of most prior
  implicit-complexity-based formalisms.
  The paper's main work is in extending and simplifying the
  original time-complexity semantics for $\ATR$ to develop
  a set of tools for extracting and solving the higher-type
  recurrences arising from feasible affine recursions.
\end{abstract}

\section{Two algorithms in search of a type-system}

As \citet{Hofmann:non-size-incr} has noted, a problem with 
implicit characterizations of complexity classes is that they
often fail to capture many natural \emph{algorithms}---usually
because the complexity-theoretic types used to control primitive
recursion impose draconian
restrictions on programming.  Here is an example.  In Bellantoni and
Cook's \citep{Bellantoni-Cook:Recursion-theoretic-char} and Leivant's
\citep{Leivant:Ram-rec-I} well-known characterizations of the
poly\-nom\-ial-time computable functions, a recursively-computed value is
prohibited from driving another recursion.  But, for instance, the
recursion clause of insertion-sort has the form $\pgm{ins\_sort}(\pgm{cons}(a,
l)) = \pgm{insert}(a, \pgm{ins\_sort}(l))$, where $\pgm{insert}$ is defined by
recursion on its second argument; selection-sort presents analogous
problems.

\citet{Hofmann:non-size-incr,Hofmann:ic03} addresses this problem 
by noting that the output of a non-size-increasing program 
(such as $\pgm{ins\_sort}$) 
should be permitted to drive another recursion, as it
cannot cause the sort of complexity blow-up the B-C-L restrictions
guard against.
To incorporate such
recursions, Hofmann defines a higher-order language with typical first-order
types and a special type~$\Diamond$ through which
functions defined recursively 
must ``pay'' for any use of size-increasing constructors, in effect
guaranteeing that there is no size increase.  Through this scheme Hofmann is 
able
to implement many natural algorithms while still ensuring that any typable
program is non-size-increasing
poly\-nom\-ial-time computable (\citet{aehlig-schw:tocl02} 
sketch an extension that
captures all of polynomial-time).  

Our earlier paper~\citep{danner-royer:ats,danner-royer:ats-lmcs}, 
hereafter referred to as~\ATS,
takes a different approach to constructing a usable programming
language with guaranteed resource usage.  We introduce a type-$2$
programming formalism called $\ATR$ (for Affine Tail Recursion, which we
rechristen in this paper as Affine \emph{Tiered} Recursion) based on
$\PCF$.  
$\ATR$'s type system is motivated by the tiering and safe/normal notions of
\cite{Leivant:Ram-rec-I} and
\cite{Bellantoni-Cook:Recursion-theoretic-char} and serves to control the
size of objects.  Instead of restricting to primitive recursion, $\ATR$
has an operator for recursive definitions; affine types and explicit clocking
on the operator serve to control time.  
We give a denotational semantics to $\ATR$ types and terms in which the
size restrictions play a key part.  This allows us, for example, to
give an~$\ATR$ \emph{definition} 
of a primitive-recursion-on-notation combinator (with
appropriate types and without explicit bounding terms) that
preserves feasibility.
We also give a \emph{time-complexity semantics} and
use it to prove that each type-$2$ $\ATR$ program has a (second-order)
polynomial run-time.%
\footnote{
These kinds of results may also have applications in the type of
static analysis for time-complexity that
\citet{frederiksen-jones:recognition} investigate.}
Finally, we show that the standard type-$2$ basic feasible functionals
(an extension of polynomial-time computability to type-$2$) 
of \citet{mehlhorn:stoc74} and \citet{cook-urquhart:fca} are
$\ATR$ definable.
Moreover,
our underlying model of computation (and complexity) is just
a standard abstract machine that implements call-by-value $\PCF$.
However, $\ATR$ is still somewhat limited as its only base type is binary
words and the only recursions allowed are tail-recursions.

\paragraph{What is new in this paper.}
In this paper we extend $\ATR$ to encompass a broad class of feasible affine
recursions.  We demonstrate these extensions by giving fairly
direct and natural versions of insertion- and selection-sorts on lists.
As additional evidence of $\ATR$'s support for programming we do not
add lists as a base type, but instead show how to implement them
over $\ATR$'s base type of binary words.

The technical
core of this paper is 
a simplification and generalization of the time-complexity semantics of~\ATS.
We construct
a straightforward framework in which recursion schemes in~$\ATR$
lead to time-complexity recurrences that must be solved to show that
these schemes preserve feasibility.  
This gives a
route to follow when adding new forms of recursion to~$\ATR$. We
follow this route to show that the recursions used to implement lists
and insertion-sort are (second-order) polynomial-time bounded.
We also discuss how to extend these results to handle the recursions
present in selection-sort.
Thus along with significantly extending our existing system
to the point where many standard algorithms can be naturally
expressed, we also provide a set of basic tools for further extensions.

\section{Programming in~$\ATR$}

\paragraph{The~$\ATR$ formalism.}
An $\ATR$ base type has the form $\Nat_L$, where \emph{labels}
$L$ are elements of the set
$(\Box\dmnd)^*\bigcup \dmnd(\Box\dmnd)^*$ (our use of~$\dmnd$
is not directly related to Hofmann's).
The labels are ordered by
$\eps\leq\dmnd\leq\Box\dmnd\leq\dmnd\Box\dmnd\leq\dotsb$
We define a subtype relation on the base types by
$\Nat_L\subtype\Nat_{L'}$ if $L\leq L'$
and extend it to function
types in the standard way.
Roughly, we can think of type-$\Nat_\varepsilon$ values as 
basic string inputs, type-$\Nat_\dmnd$ values as the result
of poly\-nom\-ial-time computations over $\Nat_\varepsilon$-values,
type-$\Nat_{\Box\dmnd}$-values as the result applying an oracle
(a type-1 input) to $\Nat_\dmnd$-values, type-$\Nat_{\dmnd\Box\dmnd}$ 
values as the result of poly\-nom\-ial-time computations over 
$\Nat_{\Box\dmnd}$-values, etc.
$\Nat_L$ is called an \emph{oracular} (respectively,
\emph{computational}) type when $L\in (\Box\dmnd)^*$ 
(respectively, $\dmnd(\Box\dmnd)^*$). 
We let $\base b$ (possibly decorated) range over base types.
Function types are formed as usual from the base types.  

The base datatype is $K=\set{\bz,\bone}^*$,
and the $\ATR$ terms are defined in Figure~\ref{fig:expr}.
The term forming operations correspond
to adding and deleting a left-most bit ($\cons_0$, $\cons_1$, and $\destr$),
testing whether a word begins with a $\bz$ or a $\bone$ ($\test_0$ and $\test_1$),
and a conditional.
The intended interpretation of $\down s\,t$ is
$s$ if $\lh s\leq\lh t$ and $\eps$ otherwise.  The recursion
operator is $\crec$, standing for clocked recursion.
\begin{figure}[tb]
\begin{align*}
s, t &::= V \mid K \mid O \mid (\lambda V.s) \mid (st) \\
  &\qquad\mid (\cons_a s) \mid (\destr s) \mid (\test_a s) \mid (\cond{s}{t_0}{t_1}) \mid (\down s\,t) \mid (\crec K(\afflambda f.t))
\end{align*}
\caption{$\ATR$ expressions.  $V$ is a set of variable symbols and
$O$ a set of oracle symbols.\label{fig:expr}}
\end{figure}

The typing rules are given in Figure~\ref{fig:typing}.
Type contexts are split (after
Barber and Plotkin's DILL~\citep{barber:dill}) into intuitionistic and
affine zones.
Variables in the former correspond 
to the usual $\arrow$ introduction and elimination rules and variables
in the latter are intended to be recursively defined; variables that occur
in the affine zone are said to \emph{occur affinely} in the term. The
\InfRule{$\crec$-I} rule 
serves as both introduction and elimination
rule for the implicit $\lollipop$~types (in the rule 
$\vec{\base b} = \base b_1,\dots,\base b_k$ and
$\vec v\oftype\vec{\base b}$ stands for $v_1\oftype \base b_1,\dots,v_k\oftype \base b_k$).
We use $\afflambda$ as the abstraction operator for variables introduced
from the affine zone of the type context to further distinguish them from
``ordinary'' variables.
The side-conditions on \InfRule{$\crec$-I} are that
$f$ occurs in cons-tail position\footnote{Informally,
$f$ occurs in \emph{cons-tail position in~$t$} if in the parse-tree
of~$t$ a path from the root to a complete application of~$f$ passes
through only
conditional branches (not tests), $\cons_0$, $\cons_1$, and the left-argument
of $\down$; $\taillh(f,t)$ is defined
to be the maximum number of $\cons_a$ operations not below any $\down$
node in any such path.\label{footnote:cons-tail-recn}} in~$t$ and
if $\base b_i\subtype\base b_1$ 
then $\base b_i$ is oracular (including $i=0$).  
The constraint on the types allows
us to prove a polynomial size-bound on the growth of the arguments to~$f$,
which in turn allows us to prove such bounds on all terms.
The typing rules enforce a ``one-use'' restriction on affine variables
by disallowing their occurrence as a free variable
in both arguments of~$\down$, the
argument of an application, the test of a conditional, or anywhere in
a $\crec$-term.

The intuition behind the \emph{shifts-to} relation $\shiftsto$ between
types is as follows.  Suppose $f\oftype \Nat_{\eps}\arrow\Nat_\dmnd$.
We think of $f$ as being a function that does some polynomial-time
computation to its input.  If we have an input~$x$ of type~$\Nat_{\Box\dmnd}$
then recalling the intuition behind the base types, we should be able
to assign the type~$\Nat_{\dmnd\Box\dmnd}$ to~$f(x)$.  The 
shifts-to relation allows us to shift input types in this way, with
a corresponding shift in output type.  As a concrete example,
the judgment $\typing{f\oftype\Nat_\eps\arrow\Nat_\dmnd,x\oftype\Nat_\eps}{}
{f(fx)}{\Nat_{\dmnd\Box\dmnd}}$ is derivable using \InfRule{Subsumption} to
coerce the type of~$f(x)$ to $\Nat_{\Box\dmnd}$ and 
\InfRule{Shift} to shift the type of the outer application of~$f$.
The definition of~$\shiftsto$ must take into account multiple
arguments and level-$2$ types and hence is somewhat involved.  Since
we do not need it for the typings in this paper, we direct the
reader to~\ATS\ for the full definition.
\begin{figure}[tb]
\begin{gather*}
\AXC{}
\LeftLabelBold{Zero-I}
\UIC{$\GDtyping\eps{\Nat_\eps}$}
\DisplayProof
\qquad
\AXC{}
\LeftLabelBold{Const-I}
\UIC{$\GDtyping K {\Nat_{\dmnd}}$}
\DisplayProof
\\
\AXC{}
\LeftLabelBold{Int-Id-I}
\UIC{$\typing{\Gamma,v\oftype\sigma}\Delta v\sigma$}
\DisplayProof
\qquad
\AXC{}
\LeftLabelBold{Aff-Id-I}
\UIC{$\typing\Gamma{\Delta,v\oftype\sigma}v\sigma$}
\DisplayProof
\\
\AXC{$\GDtyping s\sigma$}
\LeftLabelBold{Shift}
\RightLabel{($\sigma\shiftsto\tau$)}
\UIC{$\GDtyping s\tau$}
\DisplayProof
\qquad
\AXC{$\GDtyping s\sigma$}
\LeftLabelBold{Subsumption}
\RightLabel{($\sigma\subtype\tau$)}
\UIC{$\GDtyping s\tau$}
\DisplayProof
\\
\AXC{$\GDtyping s{\Nat_{\dmnd_d}}$}
\LeftLabelBold{$\cons_a$-I}
\UIC{$\GDtyping {(\cons_a s)} {\Nat_{\dmnd_d}}$}
\DisplayProof
\qquad
\AXC{$\GDtyping s {\Nat_L}$}
\LeftLabelBold{$\destr$-I}
\UIC{$\GDtyping {\destr s} {\Nat_L}$}
\DisplayProof
\qquad
\AXC{$\GDtyping s {\Nat_L}$}
\LeftLabelBold{$\test_a$-I}
\UIC{$\GDtyping {\test_a s} {\Nat_L}$}
\DisplayProof
\\
\AXC{$\typing\Gamma{\Delta_0} s{\Nat_{L_0}}$}
\AXC{$\typing\Gamma{\Delta_1} t{\Nat_{L_1}}$}
\LeftLabelBold{$\down$-I}
\BIC{$\typing\Gamma{\Delta_0,\Delta_1}{(\down st)}{\Nat_{L_1}}$}
\DisplayProof
\\
\AXC{$\Gtyping s{\Nat_{L}}$}
\AXC{$\typing\Gamma{\Delta_0} {t_0}{\Nat_{L'}}$}
\AXC{$\typing\Gamma{\Delta_1} {t_1}{\Nat_{L'}}$}
\LeftLabelBold{$\comb{if}$-I}
\TIC{$\typing\Gamma{\Delta_0\union\Delta_1}{(\cond s{t_0}{t_1})}{\Nat_{L'}}$}
\DisplayProof
\\
\AXC{$\typing{\underline{~}}{\underline{~}}K{\Nat_\dmnd}$}
\AXC{$\typing{\Gamma,\vec v\oftype\vec{\base b}}{f\oftype\vec{\base b}\arrow\base b_0}t{\base b_0}$}
\LeftLabelBold{$\crec$-I}
\BIC{$\Gtyping{\stdcrec}{\vec{\base b}\arrow\base b_0}$}
\DisplayProof
\\
\AXC{$\typing{\Gamma,v\oftype\sigma}\Delta t\tau$}
\LeftLabelBold{$\arrow$-I}
\UIC{$\GDtyping {(\lambda v.t)}{\sigma\arrow\tau}$}
\DisplayProof
\qquad
\AXC{$\GDtyping s{\sigma\arrow\tau}$}
\AXC{$\Gtyping t\sigma$}
\LeftLabelBold{$\arrow$-E}
\BIC{$\GDtyping {(st)}\tau$}
\DisplayProof
\end{gather*}
\caption{$\ATR$ typing.
The changes from~\ATS\ are
as follows: (1)~\ATS\ imposed no constraint on~$\base b_0$ 
in~(\textbf{$\crec$-I});%
(2)~\ATS\ restricted~(\textbf{$\crec$-I}) to
tail-recursion; and (3)~\ATS\ restricted~(\textbf{$\destr$-I})
and~(\textbf{$\test_a$-I}) to computational types.\label{fig:typing}}
\end{figure}

Motivated by the approach of \citet{jones:life-wout-cons}, we define
the cost of evaluation to be the size of a call-by-value evaluation
derivation.  This is essentially equivalent to the abstract machine-based
cost model of~\ATS,
but the derivation-based
model helps avoid considerable bookkeeping clutter.
Values are string constants, oracles, or abstractions.  Environments
map term variables to values or to closures over
$\crec$ terms.  A closure~$\cl t\rho$ consists of a term~$t$ and an
environment~$\rho$.
The evaluation
relation has the form~$\cl t\rho\evalto\cl z\theta$ where $\cl t\rho$ 
and $\cl z\theta$ are
closures and $z$ is a value.  The derivation rules for the evaluation
are mostly straightforward and mimic the action of the abstract
machine of~\ATS; for example, we have
\[
\AXC{$\rho(x)\evalto\cl z\theta$}
\UIC{$\cl x\rho\evalto\cl z\theta$}
\DisplayProof
\qquad
\AXC{$\cl t\rho\evalto \cl*{\bz z}\theta$}
\UIC{$\cl*{\destr t}{\rho}\evalto \cl z\theta$}
\DisplayProof
\qquad
\AXC{$\cl s\rho\evalto \cl w\zeta \qquad \cl t\rho\evalto \cl z\theta \qquad \lh w\leq\lh z$}
\UIC{$\cl*{\down st}{\rho}\evalto \cl w\zeta$}
\DisplayProof.
\]
The evaluation rule for~$\crec$ terms is
\begin{prooftree}
\AXC{}
\UIC{$\cl*{\crec a(\afflambda f.\lambda\vec v.t)}\rho\evalto \cl*{\lambda\vec v.\cond{\lh a<\lh{v_1}}{t}{\eps}}{\extend\rho f{\crec(\bz a)(\afflambda f.\lambda\vec v.t)}}$}
\end{prooftree}
which shows how unwinding the recursion increments the clock by one
step.
The cost of most inference rules is~$1$, except the $\down s\,t$ inference rules
have cost $2\lh z+1$ where $\cl t\rho\evalto \cl z\theta$
and environment and oracle evaluation have length-cost
(so, e.g., the cost of the environment rule shown above is
$\max(\lh z,1)$ when $z$ is of base type, $1$ otherwise).

\paragraph{Implementing lists and sorting.}

We implement lists of binary words via 
concatenated self-delimiting strings.  Specifically,
we code the word $w=b_0\dots b_{k-1}$ 
as $s(w) = 1b_01b_1\dots1b_{k-1}0$ and the list
$\langle w_0,\dots,w_{k-1}\rangle$
as $s(w_0)\cat\dots\cat s(w_{k-1})$, where $\cat$ is the concatenation
operation.  Code for the basic list operations is given in
Figure~\ref{fig:list-ops}.\footnote{ In these code samples, 
\lstinline[basicstyle=\footnotesize]!letrec f=s in t end! abbreviates
$\pgm{t}[f\mapsto\crec\eps(\afflambda f.s)]$ and we use the ML
notation \lstinline[basicstyle=\footnotesize]!fn x$\; \Rightarrow\dotsc$!
for $\lambda$-abstraction.}
Note that the $\pgm{cons}$, $\pgm{head}$, and $\pgm{tail}$
programs all use cons-tail recursion. Insertion-sort is expressed in
essentially its standard form, as in Figure~\ref{fig:insert-sort}.
This implementation requires another form of recursion, in
which the complete application of the recursively-defined
function appears in an argument to some operator.  In the later
part of Section~\ref{sec:rec-in-arg} we show how this
\emph{recursion in an argument} can be incorporated into $\ATR$.
Selection-sort requires yet another form of recursion
(a generalization of cons-tail recursion); we discuss how
to incorporate it into $\ATR$ in Section~\ref{sec:concl}.
\begin{figure}[tb]
\begin{lstlisting}
val nil = $\eps$ : $\Nat_{\eps}$

val cons : $\Nat_{\eps}\arrow\Nat_{\dmnd}\arrow\Nat_{\dmnd}$ = 
    fn w l $\Rightarrow$ letrec enc : $\Nat_{\eps}\arrow\Nat_{\dmnd}\arrow\Nat_{\dmnd}$ =
        fn b x $\Rightarrow$ if x then if t0(x) then c1(c0(enc b (d x)))
                            else c1(c1(enc b (d x)))
                  else c0(l)
    in enc w w end

val head : $\Nat_{\eps}\arrow\Nat_{\eps}$ =
    fn l $\Rightarrow$ letrec dec : $\Nat_{\eps}\arrow\Nat_{\dmnd}\arrow\Nat_{\dmnd}$ =
        fn b x $\Rightarrow$ if t1(x) then
                      if t0(d x) then c0(dec b (d(d(x)))) else c1(dec b (d(d(x))))
                  else $\eps$
    in down (dec l l)(l) end

val tail : $\Nat_{\eps}\arrow\Nat_{\eps}$ =
    fn l $\Rightarrow$ letrec tail' : $\Nat_{\eps}\arrow\Nat_{\eps}\arrow\Nat_{\eps}$ =
        fn b x $\Rightarrow$ if t1(x) then tail' b d(d(x)) else d(x)
    in tail' l l end
\end{lstlisting}
\caption{The basic list operations in~$\ATR$.\label{fig:list-ops}}
\end{figure}
\begin{figure}[tb]
\begin{lstlisting}
val insert : $\Nat_{\eps}\arrow\Nat_{\eps}\arrow\Nat_{\dmnd}$ =
    fn w l $\Rightarrow$ letrec ins : $\Nat_{\eps}\arrow\Nat_{\eps}\arrow\Nat_{\dmnd}$ =
        fn b l' $\Rightarrow$ if l' then
                       if leq w head(l') then cons w l'
                       else cons (head l') (ins b (tail l'))
                   else cons w nil
    in ins l l end

val ins_sort : $\Nat_{\eps}\arrow\Nat_{\dmnd}$ =
    fn l $\Rightarrow$ letrec isort : $\Nat_{\eps}\arrow\Nat_{\eps}\arrow\Nat_{\dmnd}$ =
        fn b l' = if l' then insert (head l') (down (isort b (tail l')) l') else $\eps$
    in isort l l end
\end{lstlisting}
\caption{Insertion-sort in~$\ATR$.\label{fig:insert-sort}}
\end{figure}

Our  $\pgm{head}$ and $\pgm{ins\_sort}$
programs  use the $\down$ operator to coerce the type~$\Nat_{\dmnd}$
to~$\Nat_{\eps}$.  Roughly, $\down$ is used in places
where our type-system is not clever enough to prove that the result of a
recursion is of size no larger than one of the recursion's initial
arguments; the burden of supplying these proofs is shifted off to the
correctness argument for the recursion.   A cleverer type system (say,
along the lines of Hofmann's \citep{Hofmann:ic03}) 
could obviate many of these $\down$'s,
but at the price of more complex syntax (i.e., typing), semantics (of
values and of time-complexities), and, perhaps, pragmatics (i.e.,
programming).  Our use of $\down$ gives us a more primitive (and
intensional) system than found in pure implicit complexity,%
\footnote{Leivant's \emph{recursion under a
high-tier bound}~\citep[\S3.1]{Leivant:Ram-rec-I} implements a similar idea.}
but it
also gives us a less cluttered setting to work out the basics of
complexity-theoretic compositional semantics---the focus of the rest
of the paper.  Also, in practice the proofs that the uses of $\down$
forces into the correctness argument are for the most part obvious,
and thus not a large burden on the programmer.

\section{Soundness theorems}

In this section we rework the Soundness Theorem of~\ATS\ to set up
the framework for such theorems, and then use the framework
to handle the recursions used to implement insertion-sort (we discuss
selection-sort in Section~\ref{sec:concl}).
Because of space considerations, we just sketch the main points here
and leave detailed proofs to the full paper.
The key technical notion is that of \emph{bounding} a closure~$\cl t\rho$
by a \emph{time-complexity}, which
provides upper bounds on the cost of evaluating 
$\cl t\rho$ to a value~$\cl z\theta$ as well as the 
\emph{potential} cost of using~$\cl z\theta$.  The potential of
a base-type closure is just its (denotation's) length, whereas the
potential of a function~$f$ is a function that maps potentials~$p$ to
the time complexity of evaluating~$f$ on arguments of potential~$p$.
The bounding relation gives a
\emph{time-complexity semantics} for~$\ATR$-terms; a \emph{soundness theorem}
asserts the existence of a bounding time-complexity for every $\ATR$~term.
In this paper, our soundness theorems also assert that the bounding
time-complexities are \emph{safe},
which in particular implies type-2 polynomial size and cost bounds
for the closure.  We thereby encapsulate the Soundness,
polynomial-size-boundedness, and polynomial-time-boundedness theorems of~\ATS\
(the \emph{value semantics} for the meaning of~$\ATR$ terms
and corresponding soundness theorem are
unchanged).

\paragraph{Soundness for tail-recursion.}
\label{sec:soundness}

We start by defining \emph{cost}, \emph{potential},
and \emph{time-complexity} types, all of which are elements of the
simple product type structure over the \emph{time-complexity base types}
$\set{\Tally}\union\setst{\Tally_L}{\text{$L$ is a label}}$ (we sometimes
conflate the syntactic types with their intended meaning, which is the
standard set-theoretic semantics when all base types are interpreted
as unary numerals).  The subtype relation on base types is defined by
$\Tally_L\subtype\Tally_{L'}$ if $L\leq L'$ and $\Tally_L\subtype\Tally$
for all~$L$, and extended to product and function types in the standard
way.  The
only cost type is $\Tally$, and for each $\ATR$-type~$\sigma$ we
define the potential type~$\potden\sigma$ and time-complexity 
type~$\tcden\sigma$ by
$\potden{\Nat_L} = \Tally_L$,
$\potden{\sigma\arrow\tau} = \potden\sigma\arrow\tcden\tau$, and
$\tcden\tau = \Tally\cross\potden\tau$.  
Write $\cost(\cdot)$ and $\pot(\cdot)$ for the left- and right-projections
on~$\tcden\tau$.
We introduce \emph{time-complexity variables}, a new syntactic category,
and define a time-complexity context to be a finite map from t.c.\ variables
to cost and potential types.
For a t.c.\ context~$\Sigma$, $\Env\Sigma$ is the set of
$\Sigma$ environments, defined in the usual way.
We extend $\tcden\cdot$
to $\ATR$-type contexts by introducing t.c.\ variables~$x_c$ and~$x_p$ for
each $\ATR$-variable~$x$ and setting
$\tcden\Gamma = \union_{(x\oftype\sigma)\in\Gamma}\set{x_c\oftype\Tally,x_p\oftype\potden\sigma}$.
A \emph{time-complexity denotation} of 
t.c.\ type~$\gamma$ w.r.t.\ a t.c.\ environment $\Sigma$ is a function
$X:\Env{\Sigma}\to\gamma$.
The projections $\cost$ and $\pot$ extend to t.c.\ denotations in the obvious
way.

\begin{defn} ~
\begin{enumerate}
\item 
Suppose $\cl t\rho$ is a closure and $\cl z\theta$ a value, both
of type~$\tau$; $\chi$ a time-complexity of type~$\tcden\tau$; and
$q$ a potential of type~$\potden\tau$.  Define the
\emph{bounding relations}
$\cl t\rho\apprby^\tau\chi$ and
$\cl z\theta\apprbypot^\tau q$ as follows:\footnote{We will drop the superscript
when it is clear from context.}
\begin{enumerate}
\item $\cl t\rho\apprby^\tau\chi$ if $\cost(\cl t\rho)\leq\cost(\chi)$
    and if $\cl t\rho\evalto\cl z\theta$, then 
    $\cl z\theta\apprbypot^\tau\pot(\chi)$.
\item $\cl z\theta\apprbypot^{\base b} q$ if $\lh z\leq q$.
\item $\cl*{\lambda v.t}{\theta}\apprbypot^{\sigma\arrow\tau}q$ if
    for all values~$\cl z\eta$, if $\cl z\eta\apprbypot^{\sigma} p$,
    then $\cl{t}{\extend\theta v{\cl z\eta}}\apprby^\tau q(p)$.
\item $\cl O\theta\apprbypot^{\sigma\arrow\tau}q$ if for
    all values~$\cl z\eta$, if $\cl z\eta\apprbypot^{\sigma} p$, then
    $\cl*{O(\cl z\eta)}\emptyenv\apprby^\tau q(p)$.
\end{enumerate}
\item For $\rho\in\Env{\Gamma}$ and $\varrho\in\Env{\tcden\Gamma}$, we
write $\rho\apprby\varrho$ if for all $v\in\dom\rho$ we have that
$\cl v\rho\apprby(\varrho(v_c),\varrho(v_p))$.
\item For an $\ATR$-term $\GDtyping t\tau$ and a time-complexity
denotation $X$ of type~$\tcden\tau$ w.r.t.~$\tcden{\Gamma;\Delta}$,
we say $t\apprby X$ if for all~$\rho\in\Env{(\Gamma;\Delta)}$ and
$\varrho\in\Env{\tcden{\Gamma;\Delta}}$ such that
$\rho\apprby\varrho$ we have that $\cl t\rho\apprby X\varrho$.
\end{enumerate}
\end{defn}

We define second-order polynomial expressions of tally, potential,
and time-complexity types using the operations
$+$, $*$, and $\bmax$ (binary maximum); the typing rules are given
in Figure~\ref{fig:poly-typing}.
Of course, a polynomial
$\tctyping\Sigma p\gamma$ corresponds to a t.c.\ denotation
of type~$\gamma$ w.r.t.\ $\Sigma$ in the obvious way.  We shall
frequently write $p_p$ for $\pot(p)$.
\begin{figure}[t]
\begin{gather*}
\AXC{}
\UIC{$\Stctyping \eps {\Tally_\eps}$}
\DisplayProof
\quad
\AXC{}
\UIC{$\Stctyping {\mathbf{0}^n} {\Tally_\dmnd}$}
\DisplayProof
\quad
\AXC{}
\UIC{$\tctyping {\Sigma,x\oftype\gamma} {x} \gamma$}
\DisplayProof
\\
\AXC{$\Stctyping p \gamma$}
\RightLabel{($\gamma\shiftsto\gamma'$)}
\UIC{$\Stctyping p {\gamma'}$}
\DisplayProof
\quad
\AXC{$\Stctyping p \gamma$}
\RightLabel{($\gamma\subtype\gamma'$)}
\UIC{$\Stctyping p {\gamma'}$}
\DisplayProof
\\
\AXC{$\Stctyping p {\Tally_{\dmnd_k}}$}
\AXC{$\Stctyping q {\Tally_{\dmnd_k}}$}
\BIC{$\Stctyping {p\bullet q}{\Tally_{\dmnd_k}}$}
\DisplayProof
\quad
\AXC{$\Stctyping p {\gamma}$}
\AXC{$\Stctyping q {\gamma}$}
\BIC{$\Stctyping {p\bmax q}\gamma$}
\DisplayProof
\\
\AXC{$\tctyping{\Sigma, x\oftype\sigma}{p}\tau$}
\UIC{$\tctyping\Sigma {\lambda x.p} {\sigma\arrow\tau}$}
\DisplayProof
\quad
\AXC{$\Stctyping p {\sigma\arrow\tau}$}
\AXC{$\Stctyping q \sigma$}
\BIC{$\Stctyping {pq} {\tau}$}
\DisplayProof
\end{gather*}
\caption{Typing rules for time-complexity polynomials.
$\bullet$ is $+$ or $*$,
$\gamma$ is a t.c.\ base type.
\label{fig:poly-typing}}
\end{figure}

\begin{defn}
Let $\gamma$ be a potential type, $\base b$ a time-complexity base type, 
$p$ a potential polynomial, and suppose $\Stctyping p\gamma$.
\begin{enumerate}
\item $p$ is $\base b$-strict w.r.t.~$\Sigma$ when $\tail(\gamma)\subtype\base b$
and every unshadowed\footnote{Roughly, a free-variable occurrence is
\emph{shadowed} if it is in a subterm that does not contribute to the
size of the term; see \ATS\ for details.}
free-variable occurrence in~$p$ has a type
with tail $\strictsubtype\base b$.
\item $p$ is $\base b$-chary w.r.t.~$\Sigma$ when $\gamma=\base b$ and
$p = p_1\bmax\dots\bmax p_m$ with $m\geq 0$ where 
$p_i = (vq_1\dots q_k)$ with each $q_i$ $\base b$-strict.
\item $p$ is \emph{$\base b$-safe} w.r.t.~$\Sigma$ if:
    \begin{enumerate}
    \item $\gamma$ is a base type and $p = q\pmjb r$ where
        $q$ is $\base b$-strict and $r$ is $\base b$-chary,
        $\pmjb = \bmax$ if $\base b$ is oracular, and
        $\pmjb = +$ if $\base b$ is computational.
    \item $\gamma=\sigma\arrow(\Tally\cross\tau)$ and $\pot(pv)$ is 
        $\base b$-safe
        w.r.t.~$\Sigma,v\oftype\sigma$.
    \end{enumerate}
\item A t.c.\ polynomial $\Stctyping q{\Tally\cross\gamma}$ 
is \emph{$\base b$-safe} if $\pot(q)$ is.
\item
A t.c.\ denotation $X$ of type~$\gamma$
w.r.t.~$\Sigma$ is \emph{$\base b$-safe} 
if $X$ is bounded by a $\base b$-safe t.c.\ polynomial $\Stctyping p\gamma$.
\end{enumerate}
\end{defn}

The Soundness Theorem of~\ATS\ asserts that every tail-recursive
term is bounded by a t.c.\ denotation for which the cost component
is bounded by a type-2 polynomial in the lengths of~$t$'s free variables.
In the next subsection, we extend this to cons-tail recursion and
prove that the bounding t.c.\ denotation is in fact safe.
In particular, we also have
that the potential of $t$'s
denotation is bounded by a safe polynomial.  At base type, this latter
statement corresponds to the ``poly-max'' bounds that can be computed
for Bellantoni-Cook and Leivant-style tiered functions
(e.g., \citep[Lemma 4.1]{Bellantoni-Cook:Recursion-theoretic-char}).

\paragraph{Soundness for cons-tail-recursion.}
For the remainder of this subsection $t$ is a term such that
$f$ is in cons-tail position in~$t$ and for which
we have a 
typing~$\typing{\Gamma,\vec v\oftype\vec{\base b}}{f\oftype\vec{\base b}\arrow\base b}{t}{\base b}$.
We write $\Gamma_{\vec v}$ for
for the type context $\Gamma,\vec v\oftype\vec{\base b}$.
Define the terms $C_\ell = \crec (\bz ^\ell a)(\afflambda f.\lambda\vec v.t)$ and
$T_\ell = \cond{\lh{\bz ^\ell a}<\lh{v_1}}{t}{\eps}$
(we write $\bz ^\ell a$ for $\bz\dotsc\bz a$ with $\ell$ $\bz$'s, remembering that this
is a string constant), and for any environment~$\rho$,
set $\rho_\ell = \extend\rho f{C_\ell}$.
The main difficulty in proving soundness is constructing a bounding
t.c.\ denotation for~$\crec$ terms.
A key component in the construction
is the Affine Decomposition Theorem in
Section~14 of~\ATS, which describes how to compute the
time-complexity of a term in which $f$ occurs affinely and in
tail position.  To state it, we need some definitions.

\begin{defn}
Let $X$ and $Y$ be t.c.\ denotations of type~$\tcden{\sigma\arrow\tau}$ and
$\tcden\sigma$, respectively.
\begin{enumerate}
\item For a potential~$p\oftype\Tally_L$, $\val p = (1\bmax p,p)$;
if $p$ is of higher type, then $\val p = (1, p)$.  For
a t.c.\ environment~$\varrho$ and $\ATR$ variable~$v$ we
write $\extend\varrho v{\chi}$ for
$\extend\varrho{v_c,v_p}{\cost(\chi),\pot(\chi)}$.
\item If $Y$ is 
w.r.t.~$\tcden{\Gamma,v\oftype\sigma'}$, then
$\llambda_\star v.Y \eqdef \llambda\varrho(1, \llambda v_p.Y(\extend\varrho v {\val v_p}))$
is a t.c.\ denotation of type~$\tcden{\sigma'\arrow\sigma}$
w.r.t.~$\tcden\Gamma$ (we use $\llambda x.\dotsb$ to denote the
map $x\mapsto\dotsb$).
\item 
$X\star Y \eqdef \llambda\varrho(\cost(X\varrho)+\cost(Y\varrho)+\cost(\chi)+1,\pot(\chi))$
is a t.c.\ denotation of type~$\tcden\tau$, where
$\chi = \pot(X\varrho)(\pot(Y\varrho))$ (we write $\llambda\varrho.\dots$
for $\varrho\mapsto\dots$).
\item $\dally(\ell,X) = \llambda\varrho(\ell+\cost(X\varrho),\pot(X\varrho))$
and for $\tcden\sigma = \Tally\cross\Tally_L$,
$\pad(\ell, Y) = \llambda\varrho(\cost(Y\varrho),\ell+\pot(Y\varrho))$.
\item For $\tcden\sigma =\Tally\cross\Tally_L$ and $Z$ also a t.c.\ denotation
of type~$\tcden\sigma$,
$(Z\plusmax Y)\varrho = (\cost(Z\varrho)+\cost(Y\varrho),
\pot(Z\varrho)\bmax\pot(Y\varrho))$.
\end{enumerate}
\end{defn}

\begin{thm}[Decomposition Theorem]
\label{ats-decomp}
Suppose
$t\apprby X$ and
$Y_i$ is such
that if $ft_1\dots t_k$ is a complete application of~$f$ in~$t$, then
$t_i\apprby Y_i$.
Then
$$
  t\apprby\llambda\varrho\left(X\varrho_\eps\plusmax\pad\bigl(\taillh(f,t),\varrho f\star Y_1\varrho_\eps\star\dots\star Y_k\varrho_\eps\bigr)\right)
$$
where $\varrho_\eps = \extend\varrho f{\llambda_\star\vec v.(1, 0)}$ and
$\taillh(f, t)$ is defined in Footnote~\ref{footnote:cons-tail-recn}.
\end{thm}

Intuitively,
the cost of ``getting to'' the recursive call is covered by~$X\varrho_\eps$,
and the cost of the call itself 
by~$\varrho f\star Y_1\varrho_\eps\star\dots\star Y_k\varrho_\eps$,
taking into account
any $\cons_a$ operations after the call (this is an over-estimate if no
recursive call is made).  The potential (size in this case, since~$t$
is of base type) is either independent of any complete application of~$f$
or is equal to the size of such an application, again taking into account
later~$\cons_a$ operations.

\begin{defn}
A \emph{decomposition function} for~$t$ is a function
$d(\varrho^{\Env{\tcden{\Gamma_{\vec v}}}},\chi^{\tcden\gamma})\oftype\tcden{\base b}$
such that $t\apprby\llambda\varrho.d(\varrho_\eps,\varrho f)$ (recall that
$f$ is the affinely-restricted variable in~$t$).
\end{defn}

Recalling the evaluation rule for~$\crec$ and the definition of~$\apprby$,
we see that we must understand how the closure~$\cl{T_0}{\rho_1}$
is evaluated for appropriate~$\rho$.
It is easy to see that in such an evaluation,
the only sub-evaluations of closures over terms of the form~$T_m$
are evaluations of closures of the form
$\cl{T_m}{\extend{\rho_{m+1}}{\vec v}{\vec{\cl z\theta}}}$ for some
closures~$\cl{z_i}{\theta_i}$.  For the closure~$\cl{T_0}{\rho_1}$
we say that \emph{the clock is bounded by~$K$} if in every such
sub-evaluation we have that~$\lh{z_1}<K$.

For a decomposition function~$d$ define 
$\Phi_{d,K}(n):\Env{\tcden{\Gamma_{\vec v}}}\to\tcden{\base b}$ by
\begin{align*}
\Phi_{d,K}(0) &= \llambda\varrho.(2K+1,0) \\
\Phi_{d,K}(n+1) &= \llambda\varrho.\dally\bigl(2K+1,\;d\bigl(\varrho_\eps,\dally\bigl(2,(\llambda_\star\vec v.\Phi_{d,K}(n))\varrho\bigr)\bigr)\bmax\bigl(1,0\bigr)\bigr)
\end{align*}
We will use $\Phi_{d,K}$ to bound~$T_\ell$.

\begin{thm}[Recomposition Lemma]
\label{recomposition}
Suppose $d$ is a decomposition function for~$t$,
$\rho\in\Env{\Gamma_{\vec v}}$, 
$\varrho\in\Env{\tcden{\Gamma_{\vec v}}}$,
$\rho\apprby\varrho$, and.
that in the evaluation of $\cl{T_0}{\rho_1}$ the
clock is bounded by~$K$.  Then
$\cl{T_0}{\rho_1}\apprby\Phi_{d,K}(K-\lh a)(\extend\varrho{v_i}{\val(\varrho v_{ip})})$.
\end{thm}

The Recomposition Lemma tells us that~$\Phi_{d,K}(n)$ gives us a 
bound on the time-complexity of our recursion scheme.  What we must
do now is to ``solve'' the recurrence used to define~$\Phi$ and show
that it is polynomially-bounded.

\begin{thm}[Bounding Lemma]
\label{bounding}
Suppose that in Theorem~\ref{ats-decomp} we can assume that $X$ and each $Y_i$
are bounded by t.c.\ polynomials~$p$ and $p_i$, respectively.  Assume
further that $p$ is $\potden{\base b}$-safe and $p_i$ is 
$\potden{\base b_i}$-safe w.r.t.~$\tcden{\Gamma_{\vec v}}$.  
Then there is a $\potden{\base b}$-safe polynomial
$\tctyping{\tcden{\Gamma_{\vec v}},K\oftype\potden{\base b_1},n\oftype\potden{\base b_1}}{\phi(K,n)}{\tcden{\base b}}$
such that for all $K$ and $n$, $\Phi_{d,K}(n)\leq\phi(K,n)$.
\end{thm}
\begin{proof}
Let $d$ be the decomposition function for~$t$ given in Theorem~\ref{ats-decomp}.
Using the definition of~$d$ we can find a
$\potden{\base b}$-safe polynomial 
$\tctyping{\tcden{\Gamma_{\vec v}},K\oftype\potden{\base b_1}}{(P_0(K), P_1)}{\tcden{\base b}}$ 
and
a recursive upper bound on~$\Phi_{d,K}(n)\varrho$:
\begin{align*}
\Phi_{d,K}(0)\varrho &\leq (2K+1, 0) \\
\Phi_{d,K}(n+1)\varrho &\leq (P_0(K), P_1)\varrho \plusmax \pad(\ell,\Phi_{d,K}(n)\extend\varrho{v_i}{\val(p_{ip}\varrho)})
\end{align*}
where $\ell=\taillh(f,t)$.
An easy proof by induction shows that
$\Phi_{d,K}(n)\leq (nP_0(K)\xi^{n-1}+2K+1, n\ell+P_1\xi^{n-1})$
for $n\geq 1$, where $\xi^0 = \id$ and
$(v_{ic},v_{ip})\xi^{n+1} = \val(p_{ip}\xi^n)$.
Since $\ell\not=0$ implies~$\base b_1\strictsubtype\base b$,
$n\ell+P_1\xi^{n-1}$ is bounded by
a $\potden{\base b}$-safe polynomial provided that
$P_1\xi^{n-1}$ is $\potden{\base b}$-safe.
Since $P_1$ is $\potden{\base b}$-safe and type-correct substitution of
safe polynomials into a safe polynomial yields a safe polynomial
(shown in Section~$8$ of~\ATS), to prove the theorem
it suffices to show that $p_{ip}\xi^n$ is
a $\potden{\base b_i}$-safe polynomial for each~$i$.  The proof of this
is essentially the proofs of the One-step and $n$-step lemmas
of Section~$10$ in~\ATS\ (it is here that we use the remaining constraints on 
the types in the $\crec$ typing rule).
\end{proof}

\begin{prop}[Termination Lemma]
\label{termination}
Assume the hypotheses of Theorem~\ref{bounding} hold and that
$\rho\apprby\varrho$.
Then in the evaluation of $\cl {T_{0}}{\rho_{1}}$ the
clock is bounded by $p_{1p}\xi^1\varrho$,
where $\xi^1$ is defined
as in the proof of Theorem~\ref{bounding}.
\end{prop}
\begin{proof}
This follows from the details of the proof of Theorem~\ref{bounding}.
\end{proof}

\begin{thm}[Soundness Theorem]
\label{soundness}
For every $\ATR$ term~$\GDtyping{t}{\tau}$ there is a 
$\tail(\tcden{\tau})$-safe t.c.\ denotation
$X$ of type~$\tcden\tau$ w.r.t.\ $\tcden{\Gamma;\Delta}$ such that $t\apprby X$.
\end{thm}
\begin{proof}
The proof is by induction on terms; for non-$\crec$ terms 
it is essentially as in~\ATS.  For
$\Gtyping{\crec a(\afflambda f.\lambda\vec v.t)}{\vec{\base b}\arrow\base b}$, 
suppose $\tilde\rho\in\Env\Gamma$,
$\tilde\varrho\in\Env{\tcden{\Gamma}}$, $\rho\apprby\varrho$.  
Use the Bounding, Termination and Recomposition Lemmas to show that
$\cl*{\lambda\vec v.T_0}{\tilde\rho_1}\apprby(\llambda_\star\vec v.\phi(p_{1p}\xi^1,p_{1p}\xi^1-\lh{a}))\tilde\varrho$,
where $p_1$, $\phi$, and $\xi^n$ are as in the proof of the
Bounding Lemma.
We conclude that
$\stdcrec\apprby\dally(1,\llambda_\star\vec v.\phi(p_{1p}\xi^1,p_{1p}\xi^1-\lh{a}))$.
Since this last time-complexity
is a $\potden{\base b}$-safe polynomial, the claim is proved.
\end{proof}

\begin{cor}
If $\typing{\underline{~}}{\underline{~}}{t}{\tau}$, then
$t$ is computable in type-2 polynomial time.
\end{cor}

\paragraph{Soundness for recursion in an argument.}
\label{sec:rec-in-arg}

We now address the recursions used in insertion-sort, in which the
recursive use of the function occurs inside an argument to a
previously-defined function.  What we are really after here is
structural (primitive) recursion for \emph{defined} datatypes (such as our
defined lists).  First we adapt our \InfRule{$\arrow$-E} rule to allow
affine variables to appear in arguments to applications.  We still
require some restrictions in order to ensure a one-use property;
the following is more than sufficient for our needs:
\begin{prooftree}
\AXC{$\typing\Gamma{\Delta_0} s{\sigma\arrow\tau}$}
\AXC{$\typing\Gamma{\Delta_1} t\sigma$}
\BIC{$\typing\Gamma{\Delta_0\union\Delta_1}{st}{\tau}$}
\end{prooftree}
where at most one of $\Delta_0$ and $\Delta_1$ are non-empty,
and if $\level\sigma>0$, then $\Delta_1=\emptyset$.  Thus
an affine variable~$f$ may only occur in~$t$ if $t$ is of base type,
and may not occur simultaneously in~$s$ and~$t$.
In particular, it is
safe for $\beta$-reduction to copy a completed $f$-computation, but
not an incomplete one.
To simplify notation for the recursion present in
insertion-sort we consider the special case
in which we allow typings of the form~$(*)$ provided 
$t = \cond{s'}{s(f\vec t)}{s''}$ where $f$ is not free in~$s'$ or~$s''$
(we treat the general case in the full paper).

First we must find a decomposition function.
Assuming that $s\apprby X_s$,
$t\apprby X_t$,
and $t_i\apprby Y_i$, we can take as our decomposition function
\begin{multline*}
  d(\varrho,\chi) =
  X_t\varrho\plusmax
  \bigl(\cost\bigl(X_s\varrho\bigr)+\cost\bigl(\chi\star \vec{X\varrho}\bigr)
        + \cost\bigl(\pot(X_s\varrho)(\pot(\chi\star \vec{X\varrho}))\bigr), \\
        \pot\bigl(\pot(X_s\varrho)(\pot(\chi\star \vec{X\varrho}))\bigr)\bigr)
\end{multline*}
where we have written $\chi\star\vec{X\varrho}$ for
$\chi\star X_1\varrho\star\dots\star X_k\varrho$.
Assume the inductively-given bounding t.c.\ denotations are bounded by
safe polynomials~$p_s$, $p_t$, and $p_1,\dots,p_k$.  The Soundness
Theorem follows from the Recomposition Lemma provided we have a
polynomial bound on~$\Phi_{d,K}(n)$, so now we establish such a bound.

When $\base b$ is oracular, then
since $p_{sp}$ ($=\pot(p_s)$)
is $\potden{\base b}$-safe,
we have that $p_{sp} = \lambda z^{\potden{\base b}}.(p,q_s\bmax (r_s\bmax z))$
where $q_s$ is $\potden{\base b}$-strict and $r_s$ is 
$\potden{\base b}$-chary and does not
contain~$z$.  We can therefore find a $\potden{\base b}$-safe 
t.c.\ polynomial~$(P_0(K,z^{\potden{\base b}}), P_1)$
and derive the following recursive bound
on~$\Phi_{d,K}$ using the same conventions as in our analysis of cons-tail
recursion:
\begin{align*}
\Phi_{d,K}(0)\varrho &\leq (2K+1,0) \\
\Phi_{d,K}(n+1)\varrho &= (P_0(K,\pot(\Phi_{d,K}(n)\varrho')), P_1)
  \plusmax \Phi_{d,K}(n)\varrho'
\end{align*}
where $\varrho'=\extend\varrho{v_i}{\val(p_{ip}\varrho)}$.
It is an easy
induction to show that for $n\geq 1$
$\Phi_{d,K}(n) \leq ((n\cdot P_0(K, P_1)+2K+1)\xi^{n-1},P_1\xi^{n-1})$
and thus the Bounding and Termination Lemmas that must be proved are
exactly those of before.

When $\base b$ is computational a similar calculation yields the bounding
polynomial
$((n\cdot P_0((n-2)q_s+P_1)+2p_{1p})\xi^{n-1},
(n-1)q_s\xi^{n-2}+P_1\xi^{n-1})$ for a $\potden{\base b}$-strict
polynomial~$q_s$.

\section{Concluding remarks}
\label{sec:concl}
\suppressfloats

In~\ATS\ we introduced the formalism~$\ATR$ which captures 
the basic feasible functionals at type-level~$\leq 2$.  We have
extended the formalism with recursion schemes that allow for more
natural programming and demonstrated the new formalism by implementing
lists of binary strings and insertion-sort and showing
that the new recursion schemes do not take us out of the realm of feasibility.
We have also given a strategy for proving that particular
forms of recursion can be ``safely'' added to the base system.  Here
we indicate some future directions:

\paragraph{More general affine recursions.}
In the full paper we give a definition
of \emph{plain affine recursion} that
generalizes cons-tail recursion, allows recursive
calls in arguments, and permits recursive calls in the body of
$\comb{let}$-expressions.  In particular, it covers all forms of recursion
used in the list operations and insertion- and selection-sort
(code for the latter is in Figure~\ref{fig:sel-sort}).
At the time of writing, we do not have all the details
of the soundness argument in the general case, but we expect it to
follow the framework we have developed here.  
\begin{figure}[tb]
\begin{lstlisting}
val g $\oftype\Nat_\eps\arrow\Nat_\eps\arrow\Nat_\dmnd$ =
    fn y z $\Rightarrow$ if leq y (head z) then cons y z
              else cons (head z) (cons y (tail z))

val select $\oftype\Nat_\eps\arrow\Nat_\eps$ =
    fn l $\Rightarrow$ letrec sel $\oftype \Nat_\eps\arrow\Nat_\eps\arrow\Nat_\eps$ =
        fn b l' $\Rightarrow$ if tail(l') then down (g (head l') (sel b (tail l'))) l' else l'
    in sel l l end

val sel_sort $\oftype\Nat_\eps\arrow\Nat_\dmnd$ =
    fn l $\Rightarrow$ letrec ssort $\oftype \Nat_\eps\arrow\Nat_\eps\arrow\Nat_\dmnd$ =
        fn b l' $\Rightarrow$ let val m = select l' in cons (head m) (ssort b (tail m)) end
    in ssort l l end
\end{lstlisting}
\caption{Selection-sort in~$\ATR$.  The function
\lstinline!leq! tests two integers written
in binary for inequality; we leave its full definition as an exercise for
the reader.
Note: \lstinline[basicstyle=\footnotesize]!let val x=s in t end! abbreviates
\lstinline[basicstyle=\footnotesize]!(fn x $\;\Rightarrow\;$ t)s!\label{fig:sel-sort} where we restrict~$x$ to be of base type.}
\end{figure}

\paragraph{Lazy $\ATR$.}
A version of~$\ATR$ with lazy constructors (streams) and evaluation
    would be very interesting.  There are many technical challenges in
    analyzing such a system but again we expect that the general outline
    will be the approach we have used in this paper.  
    Of course one can implement streams in
    the current call-by-value setting in standard ways (raising the
    type-level), 
    but a direct lazy implementation of streams is likely to be
    more informative.
    We expect the analysis of such a lazy-$\ATR$
    to require an extensive reworking
    of the various semantic models we have discussed here and in~\ATS.
\paragraph{Real-number algorithms.}
$\ATR$ is a type-$2$ language, but here we have focused on type-$1$
    algorithms.  We are working on implementing real-number algorithms,
    viewing a real number as a type-$1$ (stream) oracle.  This can be done in
    either a call-by-value setting  (e.g., algorithms that take a
    string of length~$n$ as input and return something like an $n$-bit
    approximation of the result) 
	or a lazy setting (in which the algorithm returns
    bits of the result on demand).

\bibliographystyle{abbrvnat}
\bibliography{master}

\appendix

\section{Typing rules and evaluation}
\label{app:typing-eval}
Recall that labels $L$ are elements
of~$(\Box\dmnd)^*\union\dmnd(\Box\dmnd)^*$.  We
define $\Box_0=\eps$, $\dmnd_d = \dmnd\Box_d$, and $\Box_{d+1}=\Box\dmnd_d$.
We give the $\ATR$ expressions and typing rules in Figures~\ref{fig:expr}
and~\ref{fig:typing}.%
\footnote{In \ATS, we restricted to tail-recursion and thus needed no
constraint on~$\base b_0$ in the (\textbf{$\crec$-I}) rule; 
we have not seen any natural programs in which this
constraint is violated.}
For convenience,
we view oracle symbols as different syntactic objects
than (type-level-$1$) variables; essentially they are variables with
a fixed meaning and that cannot be abstracted.

We define the evaluation relation
in Figure~\ref{fig:eval}.
This relates closures to values, defined simultaneously as follows:
\begin{enumerate}
\item A \emph{closure}~$\cl t\rho$ consists of a term and an environment
such that every free variable of~$t$ is in the domain of~$\rho$ and
for every~$x$ in the domain of~$\rho$, $\rho(x)$ is a closure.
\item A \emph{value}~$\cl z\theta$ is a closure in which~$z$ is either
a string constant, oracle, or abstraction.
\item An \emph{extended value}~$\cl z\theta$ is a closure that is either
a value or for which~$z = \stdcrec$ for some string constant~$a$, variables~$f$
and~$\vec v$, and term~$t$.
\item An environment is a finite map from term variables to extended values.
\end{enumerate}
Recalling that oracles range over type-$1$ functions and that the only
type-$0$ values are string constants, the evaluation rules~$O_0$ and~$O_1$
says to treat multiple-argument oracles as though they are in curried
form, returning the curried oracle result until all arguments have
been provided.  The cost of each rule is~$1$ with the following
exceptions:
\begin{enumerate}
\item The cost of (Env) is $1\bmax \lh z$ if $z$ is a string
constant and~$1$ otherwise; 
\item The cost of ($\down_i$) is $2\lh{K_t}+1$;
\item The cost of~($O_0$) is~$\lh K+1$ and the cost
of~($O_1$) is~$1$.
\end{enumerate}
These costs reflect a length-cost model of accessing the environment or
evaluating an oracle and an evaluation of $\lh{K_s}\leq\lh{K_t}$ by
stripping off bits one-by-one from each of~$K_s$ and~$K_t$.

\begin{figure}[t]
\begin{gather*}
\AXC{}
\RightLabel{($\cl z\theta$ a value)}
\UIC{$\cl z\theta\evalto\cl z\theta$}
\DisplayProof
\\
\AXC{}
\UIC{$\cl*{\stdcrec}\rho\evalto\cl*{\lambda\vec v.\cond{\lh{a}<\lh{v_1}}{t}{\eps}}{\extend\rho f{\crec(\mathbf{0}a)(\afflambda f\lambda\vec v.t)}}$}
\DisplayProof
\\
\AXC{$\rho(x)\evalto\cl z\theta$}
\LeftLabelBold{Env}
\UIC{$\cl x\rho\evalto\cl z\theta$}
\DisplayProof
\qquad
\AXC{$\cl s\rho\evalto\cl K\theta$}
\UIC{$\cl*{\cons_a s}\rho\evalto\cl*{aK}\theta$}
\DisplayProof
\\
\AXC{$\cl s\rho\evalto\cl\eps\theta$}
\UIC{$\cl*{\destr s}\rho\evalto\cl\eps\theta$}
\DisplayProof
\qquad
\AXC{$\cl s\rho\evalto\cl*{aK}\theta$}
\UIC{$\cl*{\destr s}\rho\evalto\cl K\theta$}
\DisplayProof
\qquad
\AXC{$\cl s\rho\evalto \cl*{aK}\theta$}
\UIC{$\cl*{\test_a s}\rho\evalto\cl{\mathbf{0}}\theta$}
\DisplayProof
\qquad
\AXC{$\cl s\rho\evalto\cl K\theta$}
\RightLabel{($K\not=aK'$ any~$K'$)}
\UIC{$\cl*{\test_a s}\rho\evalto\cl\eps\emptyenv$}
\DisplayProof
\\
\AXC{$\cl s\rho\evalto\cl{K_s}\theta_s$}
\AXC{$\cl t\rho\evalto\cl{K_t}\theta_t$}
\AXC{$\lh{K_s}\leq\lh{K_t}$}
\LeftLabelBold{$\down_0$}
\TIC{$\cl*{\down st}\rho\evalto \cl{K_s}\theta_s$}
\DisplayProof
\\
\AXC{$\cl s\rho\evalto\cl{K_s}\theta_s$}
\AXC{$\cl t\rho\evalto\cl{K_t}\theta_t$}
\AXC{$\lh{K_s}>\lh{K_t}$}
\LeftLabelBold{$\down_1$}
\TIC{$\cl*{\down st}\rho\evalto \cl{\eps}\emptyenv$}
\DisplayProof
\\
\AXC{$\cl s\rho\evalto\cl*{aK}\theta$}
\AXC{$\cl{t_0}\rho\evalto\cl z\theta$}
\BIC{$\cl*{\cond{s}{t_0}{t_1}}\rho\evalto \cl z\theta$}
\DisplayProof
\qquad
\AXC{$\cl s\rho\evalto\cl\eps\theta$}
\AXC{$\cl{t_1}\rho\evalto\cl z\theta$}
\BIC{$\cl*{\cond{s}{t_0}{t_1}}\rho\evalto \cl z\theta$}
\DisplayProof
\\
\AXC{$\cl s\rho\evalto\cl*{\lambda x.s'}\theta'$}
\AXC{$\cl t\rho\evalto\cl z\theta$}
\AXC{$\cl{s'}{\extend{\theta'} x{\cl z\theta}}\evalto \cl v\eta$}
\TIC{$\cl*{st}\rho\evalto \cl v\eta$}
\DisplayProof
\\
\AXC{$\cl s\rho\evalto\cl O\theta'$}
\AXC{$\cl t\rho\evalto\cl z\theta$}
\AXC{$O(\tmden z\theta) = K$}
\LeftLabelBold{$O_0$}
\TIC{$\cl*{st}\rho\evalto \cl K\emptyenv$}
\DisplayProof
\\
\AXC{$\cl s\rho\evalto\cl O\theta'$}
\AXC{$\cl t\rho\evalto\cl z\theta$}
\AXC{$O(\tmden z\theta) = O'$}
\LeftLabelBold{$O_1$}
\TIC{$\cl*{st}\rho\evalto \cl {O'}\emptyenv$}
\DisplayProof
\end{gather*}
\caption{$\ATR$ evaluation.  In the $O_i$ rules, $\tmden z\theta$ is the
denotation of~$z$ under environment~$\theta$, defined in the obvious way;
note that for a well-typed term, $z$ will be of base type, hence a
constant, so $\theta$ is irrelevant.\label{fig:eval}}
\end{figure}

The typing rules for t.c.\ polynomials are given in
Figure~\ref{fig:poly-typing}.
\begin{figure}[t]
\begin{gather*}
\AXC{}
\UIC{$\Stctyping \eps {\Tally_\eps}$}
\DisplayProof
\quad
\AXC{}
\UIC{$\Stctyping {\mathbf{0}^n} {\Tally_\dmnd}$}
\DisplayProof
\quad
\AXC{}
\UIC{$\tctyping {\Sigma,x\oftype\gamma} {x} \gamma$}
\DisplayProof
\\
\AXC{$\Stctyping p \gamma$}
\RightLabel{($\gamma\shiftsto\gamma'$)}
\UIC{$\Stctyping p {\gamma'}$}
\DisplayProof
\quad
\AXC{$\Stctyping p \gamma$}
\RightLabel{($\gamma\subtype\gamma'$)}
\UIC{$\Stctyping p {\gamma'}$}
\DisplayProof
\\
\AXC{$\Stctyping p {\Tally_{\dmnd_k}}$}
\AXC{$\Stctyping q {\Tally_{\dmnd_k}}$}
\BIC{$\Stctyping {p\bullet q}{\Tally_{\dmnd_k}}$}
\DisplayProof
\quad
\AXC{$\Stctyping p {\gamma}$}
\AXC{$\Stctyping q {\gamma}$}
\BIC{$\Stctyping {p\bmax q}\gamma$}
\DisplayProof
\\
\AXC{$\tctyping{\Sigma, x\oftype\sigma}{p}\tau$}
\UIC{$\tctyping\Sigma {\lambda x.p} {\sigma\arrow\tau}$}
\DisplayProof
\quad
\AXC{$\Stctyping p {\sigma\arrow\tau}$}
\AXC{$\Stctyping q \sigma$}
\BIC{$\Stctyping {pq} {\tau}$}
\DisplayProof
\end{gather*}
\caption{Typing rules for time-complexity polynomials.
$\bullet$ is $+$ or $*$,
$\gamma$ is a t.c.\ base type,
and $\gamma\subtype\gamma'$ is defined
by $\Tally_{\Box_k}\subtype\Tally_{\dmnd_k}\subtype\Tally_{\Box_{k+1}}$ 
and $\Tally_L\subtype\Tally$
for all~$L$.
}
\end{figure}

\section{Proofs of the main theorems}
\label{sec:proofs}

In this section, we prove the Recomposition Lemma (Theorem~\ref{recomposition}).
As a guide to the notation, environments~$\rho$ and~$\varrho$
typically refer to $\Gamma_{\vec v}$ and $\tcden{\Gamma_{\vec v}}$
environments and environments~$\tilde\rho$ and $\tilde\varrho$
typically refer to~$\Gamma$ and $\tcden\Gamma$-environments.

First we formalize the notion
of ``hard-coding'' an upper bound for the clock.  
Note that to evaluate
$\stdcrec$ applied to appropriate arguments, we really evaluate
$\cl{T_0}{\rho_1}$.
Suppose
we have a typing of the form~$(*)$ and consider the evaluation
of $\cl{T_\ell}{\rho_{\ell+1}}$  where we assume that the
$\crec$ clock-test does not terminate the recursion.  The evaluation
has the form:
\begin{prooftree}
\AXC{}
\UIC{$\cl*{0^{\ell} a}{\rho_{\ell+1}}\evalto \cl*{0^{\ell} a}{\emptyenv}$}
\UIC{$\cl*{\cons_0(0^{\ell} a)}{\rho_{\ell+1}}\evalto\cl*{0^{\ell+1}a}{\emptyenv}$}
\AXC{}
\UIC{$\cl{v_1}{\rho_{\ell+1}}\evalto\dotsb$}
\UIC{$\cl*{\cons_0(v_1)}{\rho_{\ell+1}}\evalto\dotsb$}
\BIC{$\cl*{\down(\cons_0(0^{\ell} a))(\cons_0v_1)}{\rho_{\ell+1}}\evalto\cl*{0^{\ell+1}a}\emptyenv$}
\AXC{$\mathcal D$}
\noLine
\UIC{$\cl{t}{\rho_{\ell+1}}\evalto\dotsb$}
\BIC{$\cl*{T_\ell}{\rho_{\ell+1}}\evalto\dotsb$}
\end{prooftree}
where $\mathcal D$ is the derivation
\begin{prooftree}
\AXC{}
\UIC{$\cl*{C_{\ell+1}}{\rho}\evalto\cl*{\lambda\vec v.T_{\ell+1}}{\rho_{\ell+2}}$}
\UIC{$\cl f{\rho_{\ell+1}}\evalto\cl*{\lambda\vec v.T_{\ell+1}}{\rho_{\ell+2}}$}
\UIC{$\vdots$}
\AXC{$\vdots$}
\UIC{$\cl{t_k}{\rho_{\ell+1}}\evalto\cl{z_k}{\theta_k}$}
\AXC{$\vdots$}
\UIC{$\cl*{T_{\ell+1}}{\extend{\rho_{\ell+2}}{v_i}{\cl{z_i}{\theta_i}}}\evalto\dotsb$}
\TIC{$\cl*{f\vec t}{\rho_{\ell+1}}\evalto\dotsb$}
\UIC{$\vdots$}
\UIC{$\cl{t}{\rho_{\ell+1}}\evalto\dotsb$}
\end{prooftree}
provided that $\cl t{\rho_{\ell+1}}$ actually makes a recursive call.
Thus we see that all closures over some~$T_m$ in the evaluation
of $\cl{T_\ell}{\rho_{\ell+1}}$ have the form
$\cl{T_m}{\extend{\rho_{m+1}}{\vec v}{\vec{\cl z\theta}}}$.  For a particular
closure~$\cl{T_\ell}{\rho_{\ell+1}}$ we say that the
\emph{clock is bounded by~$K$} if in its evaluation, for every subevaluation
of a closure~$\cl{T_m}{\extend{\rho_{m+1}}{\vec v}{\vec{\cl z\theta}}}$ it is
the case that $\lh{z_1}< K$.

To prove the Recomposition Lemma, we embed the evaluation of
a clocked recursion in which the clock is bounded into an evaluation
in which the clock is fixed.
To this end, introduce new term constructors $\rec_K$ with the following 
evaluation rule:
\[
  \cl*{\rec_K a(\afflambda f.\lambda\vec v.t)}{\rho}\evalto
  \cl*{\lambda\vec v.\cond{\lh a<\lh{0^K}}{t}{\eps}}{\extend\rho f{\cl*{\rec_K(0a)(\afflambda f.\lambda\vec v.t)}}}
\]
Set
\[
C_{K,\ell} = \rec_K(0^\ell a)(\afflambda f.\lambda\vec v.t)
\qquad
T_{K,\ell} = \cond{\lh{0^\ell a}<\lh{0^K}}{t}{\eps}
\]
and for an environment~$\rho$ set
$\rho_{K,\ell} = \extend\rho f{C_{K,\ell}}$.

\begin{lem}
\label{remove-lambda}
Suppose that whenever $\tilde\rho\in\Env{(\Gamma;f\oftype\gamma)}$,
$\tilde\varrho\in\Env{\tcden\Gamma}$, and
$\tilde\rho\restr\dom\Gamma\apprby\tilde\varrho$, it is the case
that $\cl*{\lambda v.t}{\tilde\rho}\apprby(\llambda_\star v.X)\tilde\varrho$.
If $\rho\in\Env{(\Gamma_{\vec v};f\oftype\gamma)}$,
$\varrho\in\Env{\tcden{\Gamma_{\vec v};f\oftype\gamma}}$, and
$\rho\restr\dom\Gamma_{\vec v}\apprby\varrho$, then
$\cl t\rho\apprby X(\extend\varrho{v_i}{\val(\varrho v_{ip})})$.
\end{lem}

\begin{defn}
For $\Phi_{d,K}$ as defined in Section~\ref{sec:soundness}, define
$\tilde\Phi_{d,K}(n) = \llambda_\star\vec v.\Phi_{d,K}(n)$.
\end{defn}

\begin{defn}
For a t.c.\ environment~$\varrho$ defined on~$\tcden{\vec v}$, define
$\varrho^V = \extend\varrho{v_i}{\val(\varrho v_{ip})}$.
\end{defn}

\begin{lem}
\label{technical-recomp}
Suppose $\Gamma,\vec v\oftype\vec{\base b};f\oftype\vec{\base b}\arrow\base b
\proves t\oftype\base b$ and that $d$ is a
decomposition function for~$t$.  
\begin{enumerate}
\item \label{item:abs}
Suppose $\tilde\rho\in\Env\Gamma$, $\tilde\varrho\in\Env{\tcden\Gamma}$,
and $\tilde\rho\apprby\tilde\varrho$.  Then
$\cl*{\lambda\vec v.T_{K,\ell}}{\tilde\rho_{K,\ell+1}}\apprby
\tilde\Phi_{d,K}(K-\lh{0^\ell a})\tilde\varrho$.
\item \label{item:base}
Suppose $\rho\in\Env{\Gamma_{\vec v}}$,
$\varrho\in\Env{\tcden{\Gamma_{\vec v}}}$, $\rho\apprby\varrho$.  
Then 
$\cl{T_{K,\ell}}{\rho_{K,\ell+1}}\apprby\Phi_{d,K}(K-\lh{0^\ell a})\varrho^V$.
\end{enumerate}
\end{lem}
\begin{proof}
The second part follows from the first by Lemma~\ref{remove-lambda}, so
we just prove the first by induction on~$K-\lh{0^\ell a}$.  The base
case is immediate.
The induction hypothesis tells us that
$\cl*{\lambda\vec v.T_{K,\ell+1}}{\tilde\rho_{K,\ell+2}}\apprby
\tilde\Phi_{d,K}(K-\lh{0^\ell a}-1)\tilde\varrho$.
Set 
$\tilde\varrho(f) = \dally(2,\tilde\Phi_{d,K}(K-\lh{0^\ell a}-1)\tilde\varrho)$.
Then since $\cl f{\rho_{K,\ell+1}}$ evaluates to
$\cl*{\lambda\vec v.T_{K,\ell+1}}{\rho_{K,\ell+2}}$ in two steps,
we have that
$\cl f{\tilde\rho_{K,\ell+1}}\apprby\dally(2,\tilde\Phi_{d,K}(K-\lh{0^\ell a}-1)\tilde\varrho) = \tilde\varrho(f)$ and
thus $\tilde\rho_{K,\ell+1}\apprby\tilde\varrho$.  
Since~$d$ is a decomposition function for~$t$, we have that
\begin{align*}
\cl*{\lambda\vec v.T_{K,\ell}}{\tilde\varrho_{K,\ell+1}}
  &= (\llambda_\star\vec v.\llambda\varrho.\dally(2K+1,d(\varrho_\eps,\varrho f)\bmax(1,0)))\tilde\varrho \\
  &= (1,\llambda v_{1p}(\dotsc(1,\llambda v_{kp}.\dally(2K+1, \\
  &\qquad d(\extend{\tilde\varrho_\eps}{v_i}{\val(v_{ip})},\dally(2,\tilde\Phi_{d,K}(K-\lh{0^\ell a}-1)\tilde\varrho)) \bmax \\
  &\qquad\qquad (1,0))) \dotsc)) \\
  &= (1,\llambda v_{1p}(\dotsc(1,\llambda v_{kp}.\dally(2K+1, \\
  &\qquad d(\extend{\tilde\varrho_\eps}{v_i}{\val(v_{ip})}, \\
  &\qquad\qquad\dally(2,\tilde\Phi_{d,K}(K-\lh{0^\ell a}-1)\extend{\tilde\varrho}{v_i}{\val(v_{ip})}))\bmax \\
  &\qquad\qquad (1,0))) \dotsc)) \\
  &=(\llambda_\star\vec v.\llambda\varrho.\dally(2K+1, \\
  &\qquad d(\varrho_\eps,\dally(2,\tilde\Phi_{d,K}(K-\lh{0^\ell a}-1)\varrho))\bmax(1,0)))\tilde\varrho \\
  &=(\llambda_\star\vec v.\Phi_{d,K}(K-\lh{0^\ell a}))\tilde\varrho \\
  &=\tilde\Phi_{d,K}(K-\lh{0^\ell a})\tilde\varrho.
\end{align*}
\end{proof}

\begin{thm}[Theorem~\ref{recomposition}: Recomposition Lemma]
Assume the hypotheses of Lemma~\ref{technical-recomp}(\ref{item:base}).
Assume further that in the evaluation of $\cl{T_0}{\rho_1}$ the
clock is bounded by~$K$.  Then
$\cl{T_0}{\rho_1}\apprby\Phi_{d,K}(K-\lh a)\varrho^V$.
\end{thm}
\begin{proof}[Proof Sketch]
The hypotheses allow us to define a injective map $F$ from the evaluation 
derivation of $\cl{T_0}{\rho_{1}}$ to the evaluation derivation of
$\cl{T_{K,0}}{\rho_{K,1}}$ such that:
\begin{enumerate}
\item $F$ maps the root to the root;
\item $F$ preserves the ``child-of'' relation;
\item The only differences between the closures at the node~$x$ and $F(x)$ are:
	\begin{enumerate}
    \item $C_{m}$ is replaced with $C_{K,m}$;
    \item $T_{m}$ is replaced with $T_{K,m}$;
    \item \sloppypar The evaluations of 
    	$\cl*{\down(\cons_0(0^{m} a)(\cons_0v_1))}{\rho'_{m+1}}$
    	are mapped to evaluations of
		$\cl*{\down(\cons_0(0^{m} a)(\cons_00^K))}{\rho'_{m+1}}$.
	\end{enumerate}
\end{enumerate}
Thus we have that the evaluation derivation of $\cl{T_0}{\rho_{1}}$ 
is no
larger than that of $\cl{T_{K,0}}{\rho_{K,1}}$ and that
$\cl{T_0}{\rho_1}\evalto \cl{z}{\theta}$ iff
$\cl{T_{K,0}}{\rho_{K,1}}\evalto\cl z\theta$.
From this we conclude that since
$\cl*{T_{K,0}}{\rho_{K,1}}\apprby\Phi_{d,K}(K-\lh{a})\varrho^V$
we also have that
$\cl*{T_0}{\rho_{1}}\apprby\Phi_{d,K}(K-\lh{a})\varrho^V$.
\end{proof}

\begin{thm}[Theorem~\ref{soundness}:  Soundness Theorem]
If $\GDtyping t\tau$ is an~$\ATR$ term, then there is a 
$\tail(\tcden\tau)$-safe t.c.\ denotation~$X$ of type~$\tcden\tau$
w.r.t.~$\tcden{\Gamma;\Delta}$ such that $t\apprby X$.
\sloppypar
\end{thm}
\begin{proof}
The proof is by induction on~$t$.  For everything but~$\crec$ terms,
it is mostly a pushing-through of the definition of~$\apprby$.  Now
suppose that $\Gtyping\stdcrec{\vec{\base b}\arrow\base b}$,
$\tilde\rho\in\Env\Gamma$, $\tilde\varrho\in\Env{\tcden\Gamma}$,
and that $\tilde\rho\apprby\tilde\varrho$.  Noting that
$\cl*{\stdcrec}{\tilde\rho}\evalto \cl*{\lambda\vec v.T_0}{\tilde\rho_1}$,
we wish to show that this latter term is bounded
by $(\llambda_\star\vec v.\phi(p_{1p}\xi^1,p_{1p}\xi^1-\lh{a},\vec v))\tilde\varrho$
where $\phi$ and $\xi$ are as in the proof of the Bounding Lemma.
To do so, it suffices to show that if $\cl{z_i}{\theta_i}\apprbypot q_i$,
$\tilde\rho_1^* = \extend{\tilde\rho_1}{v_i}{\cl{z_i}{\theta_i}}$,
and $\tilde\varrho^* = \extend{\tilde\varrho}{v_i}{\val(q_i)}$,
then $\cl{T_0}{\tilde\rho_1^*}\apprby
\phi(p_{1p}\xi^1,p_{1p}\xi^1-\lh a,\vec v)\tilde\varrho^*$.  
Since $\tilde\rho_1^*\apprby
\tilde\varrho^*$, from the Termination Lemma we have that the clock
on~$\cl{T_0}{\tilde\rho_1^*}$ is bounded by
$p_{1p}\xi^1\tilde\varrho^*$.  Thus by the Recomposition Lemma
we have that 
\begin{multline*}
\cl{T_0}{\tilde\rho_1^*}\apprby
\Phi_{d,p_{1p}\xi^1\tilde\varrho^*}(p_{1p}\xi^1\tilde\varrho^*-\lh a)(\tilde\varrho^*)^V = 
\Phi_{d,p_{1p}\xi^1\tilde\varrho^*}(p_{1p}\xi^1\tilde\varrho^*-\lh a)\tilde\varrho^* \\
\leq\phi(p_{1p}\xi^1,p_{1p}\xi^1-\lh a,\vec v)\tilde\varrho^*.
\end{multline*}
We conclude that
$\cl*{\stdcrec}{\tilde\rho}\apprby\dally(1,\llambda_\star\vec v.\phi(p_{1p}\xi^1,p_{1p}\xi^1-\lh a,\vec v))\tilde\varrho$
and hence that
$\stdcrec\apprby\dally(1,\llambda_\star\vec v,\phi(p_{1p}\xi^1,p_{1p}\xi^1-\lh a,\vec v))$.
\end{proof}

\section{Plain affine recursion}
\label{app:recn-in-argument}
We generalize the recursion schemes we have discussed in this paper as follows:
\begin{defn}
$t$ is a \emph{plain affine recursive definition of~$f$} if:\footnote{%
Clearly we are duplicating work that the affine type system does
for us here; we have not yet fully investigated this situation.}
\begin{enumerate}
\item $f\notin\fv(t)$; or
\item $t = ft_1\dots t_k$ where $f\notin\fv(t_i)$ for any~$i$;
\item $t = \cond{s}{s_0}{s_1}$ where $f\notin\fv(s)$ and 
each $s_i$ is a plain affine recursive definition of~$f$; or
\item $t=\comb{op} s$ where $\comb{op}$ is any of $\cons_a$, $\destr$,
or~$\test_a$ and $s$ is a plain affine recursive definition of~$f$; or
\item $t=\down s_0s_1$ where $s_0$ is a plain affine recursive definition
of~$f$ and
$f\notin\fv(s_1)$; or
\item $t = st_1\dots t_k$ where $f\notin\fv(s)$ and each~$t_i$ is a
plain affine recursive definition of~$f$; or
\item $t = (\lambda x.s)r$ where $s$ is a plain affine recursive
definition of~$f$ and $f\notin\fv(r)$.
\end{enumerate}
\end{defn}

We continue here to consider the special case of
$t = \cond{s'}{s(f\vec t)}{s''}$ where $f$ is not free in~$s'$ or~$s''$.
We have already established a decomposition function; all that remains
to to set up and solve a recursive bound on~$\Phi_{d,K}(n)$ when
$\base b$ is computational.
In this case
$p_{sp} = \lambda z^{\potden{\base b}}.(p, q_s+(r_s\bmax z))$ where $q_s$
is $\potden{\base b}$-strict and $r_s$ is $\potden{\base b}$-chary and
does not contain~$z$.  The recurrence to solve is
\begin{align*}
\Phi_{d,K}(0)\varrho &\leq (2K+1,0) \\
\Phi_{d,K}(n+1)\varrho &\leq (P_0(K, \pot(\Phi_{d,K}(n)\varrho')), P_1)\plusmax
  \pad(q_s,\Phi_{d,K}(n)\varrho')
\end{align*}
where $\varrho'=\extend\varrho{v_i}{\val(p_{ip}\varrho)}$
and $(P_0(K, z^{\potden{\base b}}), P_1)$ is a 
$\potden{\base b}$-safe t.c.\ polynomial.
The solution
to this recurrence is given by
\[
\Phi_{d,K}(n) \leq ((n\cdot P_0((n-2)q_s+P_1)+2K+1)\xi^{n-1},
(n-1)q_s\xi^{n-2}+P_1\xi^{n-1})
\]
for $n\geq 2$, so the Bounding and Terminations Lemmas to be proved
are those of before.  Furthermore, since $\base b$ is computational,
$\base b>\base b_1$ and so
we have that $n^{\potden{\base b_1}}q_s$ is a
$\base b$-strict polynomial, and hence
$nq_s\xi^{n-1}+P_1\xi^n$ is $\base b$-safe for each~$n$.
The rest of the Soundness Theorem follows.

\end{document}